\documentclass[%
 reprint,
superscriptaddress,
 amsmath,amssymb,
 aps,
]{revtex4-2}

\usepackage{comment} 

\usepackage{graphicx}
\usepackage{dcolumn}
\usepackage{bm}

\usepackage{subcaption} 

\newcommand{\sz}[0]{\sigma^z}
\newcommand{\sx}[0]{\sigma^x}

\newcommand{\ket}[1]{|#1\rangle}

\begin{document}

\preprint{APS/123-QED}

\title{Steered quantum annealing: \\ improving time efficiency with partial information}

\author{Ana Palacios de Luis}
\email{ana.palacios@qilimanjaro.tech}
\affiliation{%
 Qilimanjaro Quantum Tech, Barcelona 08007, Spain
}
\author{Artur Garcia-Saez}
\affiliation{%
 Qilimanjaro Quantum Tech, Barcelona 08007, Spain
}
\affiliation{Barcelona Supercomputing Center, Barcelona 08034, Spain.}
\author{Marta P. Estarellas}%
\affiliation{%
 Qilimanjaro Quantum Tech, Barcelona 08007, Spain
}

\date{\today}

\begin{abstract}
In the computational model of quantum annealing, the size of the minimum gap between the ground state and the first excited state of the system is of particular importance, since it is inversely proportional to the running time of the algorithm. Thus, it is desirable to keep the gap as large as possible during the annealing process, since it allows the computation to remain under the protection of the adiabatic theorem while staying efficient. We propose steered quantum annealing as a new method to enlarge the gap throughout the process, in the case of diagonal final Hamiltonians, based on the exploitation of some assumptions we can make about the particular problem instance. In order to introduce this information, we propose beginning the anneal from a biased Hamiltonian that incorporates reliable assumptions about the final ground state. Our simulations show that this method yields a larger average gap throughout the whole computation, which results in an increased robustness of the overall annealing process. 
\end{abstract}

\maketitle


\section{Introduction}

Quantum annealing is an analog model of quantum computation that, in line with the rest of quantum technologies, has been on the rise in the last decades \cite{albash_adiabatic_2018, crosson_prospects_2020}. While capable of being universal (under certain conditions \cite{biamonte_realizable_2008, cubitt_universal_2018}), this model is usually exploited for the purpose of classical optimisation tasks \cite{lucas_ising_2014, albash_adiabatic_2018}. Quantum annealers allow for the implementation of heuristic and exact algorithms to very hard problems. These devices may harness a potential advantage due to the presence of quantum phenomena, inaccessible to classical methods \cite{crosson_prospects_2020}. This is achieved by encoding the solution of the problem in the ground state (GS) of some Hamiltonian that can be implemented in the machine. Once the problem is formulated in this manner, it can be solved by initialising the system in a different ground state that is easy to prepare and slowly changing said system's parametrisation in order to arrive to the problem Hamiltonian \cite{farhi_quantum_2000}. The key term here is ``slowly'', which relates to the physical principle this protocol is based on: the adiabatic theorem \cite{Born_1928, albash_adiabatic_2018}. This theorem states that for a system that is in the ground state of a given Hamiltonian, if the latter suffers some small (or, equivalently, slow) change, the system will end up in the ground state of the new Hamiltonian. The total time the process must take in order to follow the prediction of the adiabatic theorem is inversely proportional to the energy gap between the ground state and the first excited state \cite{Kato_1950}. Thus, for the sake of time efficiency we are interested in large gaps in order to have shorter computation times. However, large problems present polynomially or, in the worst cases, exponentially closing gaps for increasing system sizes, which has led to the exploration and development of a series of techniques to tend to these issues. The explored approaches are very diverse: implementing non-trivial schedules that distribute the time spent in different stages of the anneal optimally (for which a canonical example can be found in \cite{roland_quantum_2001}), introducing additional terms, referred to as catalysts (e.g., \cite{farhi_quantum_2002, crosson_different_2014, albash_catalysts_2019}) and/or modifying the initial Hamiltonian \cite{farhi_quantum_2011, Perdomo-OrtizVA11}. Another approach is to abandon the adiabatic premise and try to take advantage of diabatic transitions, either by explicitly engineering against them (with the so-called counter-diabatic driving \cite{prielinger_counterdiabatic_2020}) or by other means to shortcut towards adiabaticity \cite{crosson_different_2014, guery-odelin_shortcuts_2019}.

To tackle this issue from an adiabatic quantum computing (AQC) perspective, we propose to ``steer'' the quantum annealing process by leveraging information in the form of a recommended subspace, whose nature is based on assumptions about the solution of the problem encoded in the final Hamiltonian's ground state. If the assumptions are accurate enough, we steer the annealing process towards the target state in a more efficient manner by introducing this information into the algorithm such that the relevant gap is enlarged, ultimately resulting in a faster computation. We are guiding our annealing path according to our recommendation without the need for preparing any state, as we introduce its information via a global rotation of the standard initial Hamiltonian. Importantly, with a simple parametrisation of such rotation one can control the confidence they have on the proposed recommended state. The source of the information used to construct the recommended state will generally come from the physical intuition behind the problem at hand, thus making its nature strictly problem-dependent. However, as we will illustrate later on, a reasonable guess may sometimes be extracted from the final Hamiltonian as well. Approximate solutions resulting from various classical heuristic algorithms may also provide some reliable information about the global minimum that our method can refine in order to yield the exact solution. At the time being there are very few proposals for feeding additional information to a quantum annealing algorithm, many of which are based on the idea of reverse annealing \cite{yamashiro_dynamics_2019}. The main advantage of our technique with respect to the latter is that it allows us to exploit partial information about the solution instead of a full state, a very interesting feature that, to our knowledge, is not present in any other proposal up to date. A somewhat related work to the one presented here is that in \cite{Perdomo-OrtizVA11}, where they propose starting from a Hamiltonian whose ground state corresponds to a state of the computational basis, selected by heuristics, and transition towards the problem Hamiltonian by applying a catalyst on the $x$-direction. Another relevant advantage with respect to some other techniques is that we do not require additional qubits or interaction terms, which allows for a simple implementation. On the other hand, in the context of QAOA, a recent work has explored a similar method to the one proposed here to introduce additional information (in this case, coming from heuristic approaches) into a parametrised circuit \cite{Egger2021warmstarting}.


\section{Methodology}

In this section we will explore the specifics of our proposed technique. In AQC, the usual initial Hamiltonian of the anneal is $H_0 = -\sum_{i}^N \sx_i$, whose ground state is the equal superposition of all the $2^N$ elements of the computational basis $\{\ket{\phi_j}\}$:
\begin{gather}
    \ket{+} = \frac{1}{\sqrt{2^N}} \sum_{j}^{2^N} \ket{\phi_j}
\end{gather}
The standard annealing process consists on interpolating linearly between this initial Hamiltonian $H_0$ and the final Hamiltonian $H_f$ that encodes the solution.
\begin{gather}
    H(s) = (1-s) H_0 + s H_f
\end{gather}
where $s$ is an adimensional time $s=\frac{t}{T} \in [0, 1]$ that parametrises the progression of the anneal and $T$ is the duration of the full process in real time.

Our scheme consists on starting from a unitary transformation of this $H_0$, $\Tilde{H}_0$, that encodes some assumption we can make about our ground state (GS), such that the overlap of the new initial GS with the sought-after solution is larger. The expression of this new initial Hamiltonian $\Tilde{H}_0$ is
\begin{gather}
    \Tilde{H}_0 = R_y^{\dagger}(\vec{\theta}) H_0 R_y(\vec{\theta}) = \sum_i^N -(\cos{(\theta_i)} \sx_i + \sin{(\theta_i)} \sz_i)
    \label{H_rot}
\end{gather}
with $\vec{\theta} = \Theta \cdot \text{sgn}[\vec{\psi}]$. We assume $\hbar=1$ throughout the paper. $\Theta$ is an angle of our choice and $\vec{\psi}$ is the vector containing the initially guessed information, whose elements can take the values $\pm 1$ or $0$.  For example, in case we have reason to believe that the third and fourth qubits (i.e., spins in our context) are pointing upwards and downwards respectively in the solution, we would have $\vec{\psi} = (0, 0, +1, -1, 0, ..., 0)$, such that we are only introducing a bias in these two qubits. We may also highlight that when $\Theta = 0$, Eq. \eqref{H_rot} reduces to a canonical annealing process, which we will refer to as a direct anneal. It is important to note that all the terms present in $\Tilde{H}_0$, in the case of superconducting qubits, can be implemented in a similar way to those required to perform the direct anneal. With this new initial Hamiltonian, the resulting annealing process is now described as
\begin{gather}
    H(s) = (1-s) \Tilde{H}_0 + s H_f
    \label{justacatalyst}
\end{gather}

\begin{figure}
    \includegraphics[width = 0.35\textwidth]{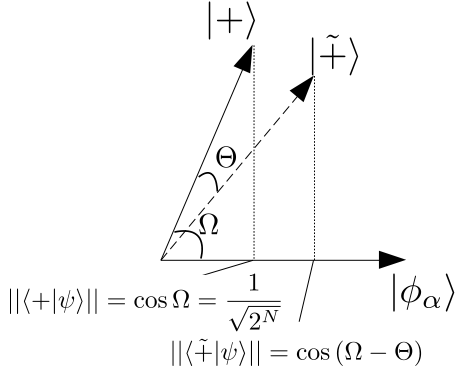}
    \caption{Illustration of the idea behind the rotation of $H_0$. For the sake of simplicity let us assume here that the final ground state $\ket{\phi_{\alpha}}$ is known and therefore fully introduced as the initial guess $\vec{\psi}$. $\ket{\Tilde{+}}$ is the ground state of $\Tilde{H}_0$.}
    \label{fig:rot_hi}
\end{figure}

Let us now move to a more intuitive picture on how this method works. To get an illustration on the relevance of the parameter $\Theta$, we consider the simple scenario of state preparation: in this case, we know the state we want to prepare, say $\ket{\phi_{\alpha}}$, and therefore the true solution beforehand. At time $t =0$, the overlap of the initial GS, $\ket{+} = \frac{1}{\sqrt{2^N}} \sum_j \ket{\phi_j}$, with the final solution will be $|| \langle + | \phi_{\alpha} \rangle ||= \frac{1}{\sqrt{2^N}} = \cos{\Omega}$. As shown in Fig.~\ref{fig:rot_hi}$, \Omega$ can then be interpreted as the angle between the  initial and final ground states. The role of $\Theta$ in \eqref{H_rot} is thus evidenced to be the rotation angle applied to the initial ground state, $\ket{+}$, such that the GS of $\Tilde{H}_0$, $\ket{\widetilde{+}}$ (see its precise expression Eq. \eqref{unperturbedGS_Hm} in the Appendix) now has a bigger overlap with $\ket{\phi_{\alpha}}$. In short, by aiming at a subspace that contains the final solution we expect the ground state of our system to be closer to the solution throughout the whole anneal, thus producing a lighter computation. We also note that, since we are also capable of bringing the state further away from a certain subspace by setting $\Theta < 0$, we may use this procedure to explicitly disrecommend states that have been identified as local minima in the search for better solutions.

In spite of the simplification contained in Fig.~\ref{fig:rot_hi}, this illustration already suggests that $\Theta$ should be looked at in units of $\Omega$, which is a function of $N$ and
therefore allows us to extrapolate this normalisation to larger systems.


\subsection{\label{sec:pert_GS_Hm} Ground states of $H(s)$ for low $s$} 

To provide a more detailed insight of our method we take a closer look into the ground state of $H(s)$ at the initial stages of the anneal, i.e, for low $s$. For this, we derive the perturbed GS wave function of $H(s)$ up to second order around $\Tilde{H}_0$, which is presented and analysed in further detail in Appendix \ref{sec:pert_appendix}.

Thus, we take $\Tilde{H}_0$ as the unperturbed Hamiltonian and perturb with $H_f$ according to the point of the anneal we are examining, $s^*$. In this manner, we may write the examined Hamiltonian as follows:
\begin{gather}
    H = \Tilde{H}_0 + \varepsilon H_f \\
    \varepsilon = \frac{s^*}{1-s^*} 
\end{gather}

\begin{figure}[h]
    \includegraphics[width = 0.49\textwidth]{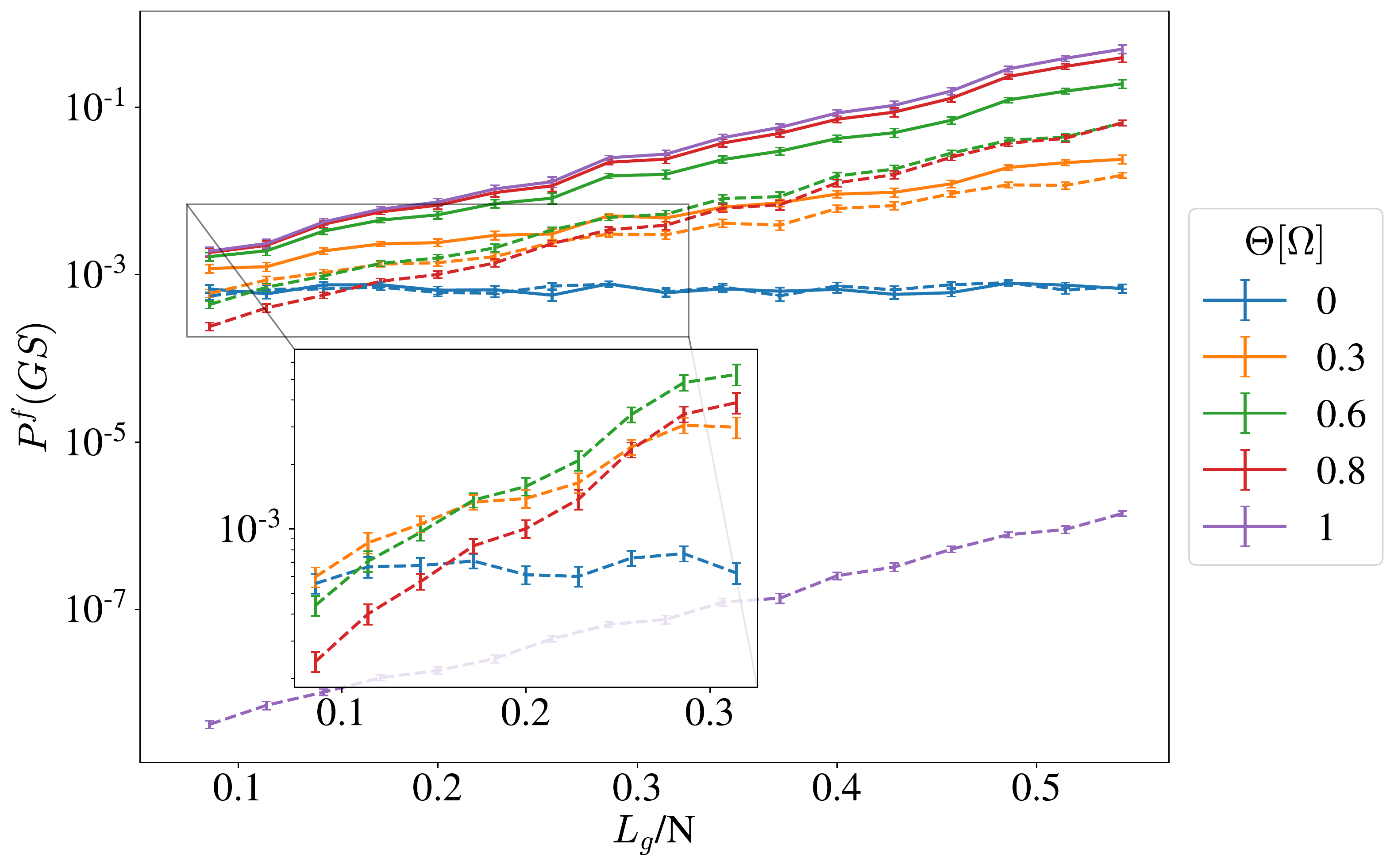}
    \caption{Probability of finding the final target state when measuring the ground state of $H$ ($s^* = 0.3$) as we increase the size of our guess, $L_g$, for $N=35$ when it is fully correct (solid lines) or contains a single wrong assignment (dashed lines). Averages are taken over 20 random instances of parameters $J_{mean} \simeq -0.007$, $\sigma_J \simeq 14$ and $h_{mean} \simeq 1.22, \sigma_h \simeq 0.08$.}
    \label{fig:pert_th_vsN}
\end{figure}

Our main goal in this section is to obtain a qualitative picture of how the nature of our initial guess $\vec{\theta}$
affects the overall procedure. For this, we analyse the overlap of the approximated ground state of $H$ with the final GS of $H_f$ for different problem instances. We note that in order to analyse systems of large sizes, which are out of reach for exact numerical methods, 
each instance is an artificial, randomly generated solution of the desired length. As it can be seen in Appendix \ref{sec:pert_appendix}, the obtained expressions to construct the analytic form of the GS provided by perturbation theory also require the inclusion of the Hamiltonian's parameters, namely $|\sum_i h_i|$ and $|\sum_{i, j > i} J_{ij}|$ in the case of an Ising model. For the numerical evaluation of these parameters in the analytic expressions we make use of the central limit theorem, which guarantees that we can sample these sums (without the absolute value) from a Gaussian distribution of certain mean values $\overline{J}, \overline{h}$ and standard deviation $\sigma_J, \sigma_h$. In order to match the Ising scenario that will be discussed further in the upcoming section, we have determined these parameters by considering the uniform distributions $J_{ij}\in[-1, 1]$ and $h_i \in [h_{mean} - W, h_{mean} + W]$ for $h_{mean} = 0.01, W = 0.05$ (and the specified system size in this case, $N=35$) and fitting the resulting Gaussians. This has resulted in $\overline{J} \simeq -0.007, \sigma_J \simeq 14$ and $\overline{h} \simeq 1.22, \sigma_h \simeq 0.08$. The results obtained according to these statistics are presented in Fig.~\ref{fig:pert_th_vsN},
which shows the probability of finding the final GS at $s^*=0.3$, $P^f(GS)= || \langle GS_{s^*}\vert GS_f\rangle ||^2$, depending on the relative size, $L_g$, of the initial guess (i.e. number of non-zero elements in $\vec{\psi}$) and the presence (dashed lines)/absence (solid lines) of a single incorrect assignment for different values of the angle $\Theta$. We observe that for $\Theta \neq 0$, $P^f(GS)$ increases exponentially with the amount of partial information introduced to the algorithm.
The inset of Fig.~\ref{fig:pert_th_vsN} highlights the benefits of setting $\Theta < \Omega$ in terms of robustness to the presence of one incorrect assignment (wrong non-zero element of $\vec{\psi}$) in our initial guess. For example, for $L_g/N < 0.2$, we find a higher overlap with the final solution for $\Theta = 0.3\Omega$ than for $\Theta = 0.8\Omega$, which actually does worse than the direct case. In Appendix \ref{sec:scaling_correctness} we present some further analysis on this direction by studying the overlap for an increasing number of correct assignments for a guess of fixed length.

The perturbative analysis of the GS of $H(s)$ for low values of $s$ therefore supports that an overall faster approach towards the final target state can be enforced from the initial stages of the anneal. In fact, this guidance is the underlying reason of the enlarged gap we observe for the entire (successful) protocol, as we will see with more detail in the next section.

%

\section{Numerical results}

Let us now move to a more exhaustive analysis of the numerical results obtained using our technique in two problem Hamiltonians common in quantum annealing: the Ising model and the QUBO (Quadratic Unconstrained Binary Optimisation) formulation of 3SAT.

\subsection{Ising model} \label{sec:Ising}
We here focus on the study of the random Ising model 
\begin{gather}
    H_f = \sum_i h_i \sz_i + 
    \sum_{j>i} J_{ij} \sz_i \sz_j
    \label{IsingHami} \\
    h_i = h_{mean} + W_i
\end{gather}
with long-range interactions $\{J_{ij} \} \in [-J_s, J_s]$ and local fluctuations $\{W_i\} \in [-W, W]$, both sampled from uniform distributions. Throughout this paper, we will fix the energy scale to  $J_s = 1$, therefore fixing our time units through its inverse as well. Different parameter regimes as a function of both $h_{mean}$ and $W$ are explored, with the greatest success of the protocol (with respect to a simple, direct anneal between $H_0$ and $H_f$) being found in the spin-glassy regime ($J \gg h$), as opposed to the high-$h$ regime. Because of that we will not cover the high local field regime ($h \gg J$) in our analysis since it corresponds to rather trivial problems, where the final ground state is mostly oblivious to the interactions present. Instead, we will focus on the spin-glass regime, a common model to formulate classical optimisation problems. We will also treat the case with $J \sim h$ in Section \ref{sec:3SAT} in the context of the Ising formulation of 3SAT problems.

In the spin-glassy regime of the Ising model, a simple analysis of its symmetry can provide us with a reasonable initial guess to test our procedure. In this case the interaction terms, $J_{ij}\sigma^z_i \sigma^z_j$, which constitute the dominant contribution to the Hamiltonian, are symmetric under spin inversion, a symmetry that is only broken by the local terms, $h_i\sigma^z_i$. Thus, if we make the assumption that the highest local field, $\max_i |h_i|$, is the main responsible for the breaking of degeneracy, we may consider that the spin with the highest local field will be pointing in the direction determined by it as our initial guess. Such educated guess turns out to be correct in about 75\% of the cases considered, and is an example of how we may extract simple partial information about our conjectured solution to be fed to the algorithm.
We numerically checked the validity of this assumption on an ensemble of 1000 random instances with $N=6$ spins, which was correct for 769 instances, and for the same number of samples with $N=8$ spins, correct for 754 instances.

Let us now explore the benefits of our method for different instances of the target Hamiltonian in \eqref{IsingHami}. For the numerical simulation of the adiabatic algorithm we have used exact diagonalisation for each step (with an adimensional simulation time step of $\delta s=\frac{T}{\delta t} = 0.01$, which will be the default value in all our experiments, where $T$ is the total time set in the simulation) using the software package Qibo \cite{efthymiou_qibo_2021}. Our aim is to have a picture of the energy landscape as well as an estimation of the overall annealing time, a quantity that is inversely proportional to the gap between the ground state and the first excited state, $\Delta$, throughout the entire annealing process. The estimated total annealing time will be extracted from the instantaneous adiabatic time, $T_{ad}(s)$, which is defined as follows:
\begin{gather}
    T_{ad}(s) = \frac{||\partial_s{H}(s)||}{\Delta(s)^2} \cdot ||H(s)||
\end{gather}
where $||A||$ refers to the L2-norm of a given matrix $A$. We note that there are several nonequivalent expressions for the adiabatic time bound depending on how it is derived (see \cite{albash_adiabatic_2018, jansen_bounds_2007} for more details). We are following the one with an inverse square dependence on gap size, which is the most widespread \cite{messiah_quantum_1962} and is in good agreement with our simulation results. We take the integration of $T_{ad}(s)$ for $s \in [0, 1]$ as the estimated total adiabatic time, which will provide us with a tighter bound than the one typically taken in other works \cite{albash_adiabatic_2018}. The instantaneous profile $T_{ad}(s)$ also provides us with information about the optimal speed at which the anneal should progress at all times, since we may obtain an optimal adiabatic schedule by inverting its cumulative as follows:
\begin{gather}
    t(s) = \int_0^s T_{ad}(s) ds \quad \Rightarrow \quad
    s(t) = [t(s)]^{-1}
    \label{svst}
\end{gather}
All in all, the instantaneous adiabatic time is a useful tool to better understand an annealing process.

\begin{figure*}
    \includegraphics[width = \textwidth]{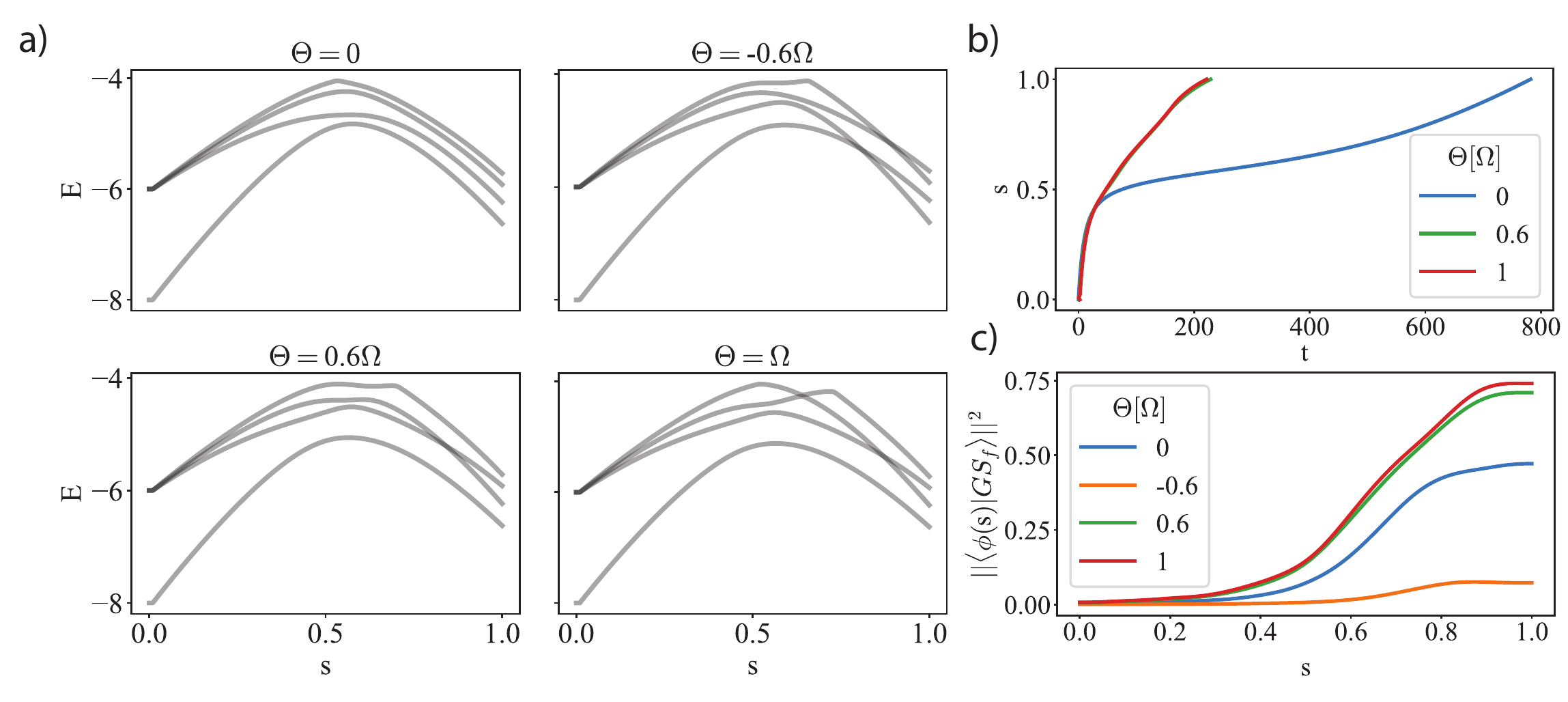}
    \caption{Results for an Ising instance with $N=8$, $h_{mean} = 0.01$, $W=0.05$ and a favourable initial guess for different values of $\Theta$. a) Energy landscape of the lowest 4 levels (grey solid line). b) Optimal adiabatic schedules resulting from the direct ($\Theta = 0$) and the relevant steered processes, $\Theta=0.6\Omega$ and $\Theta=\Omega$. c) Probability of finding the target ground state throughout the anneal, where $\ket{GS_f}$ is the final ground state and $\ket{\phi(s)}$ is the state of the system at time $s$. The total annealing time for the simulation to obtain the probabilities in c) is set to $T=15$.}
    \label{fig:E_ov_oas_rothi}
\end{figure*}

In Fig.~\ref{fig:E_ov_oas_rothi} we show the effect of the proposed method on one random Ising instance with a correct initial guess for different values of $\Theta$. We also illustrate a case with $\Theta < 0$, in which we are \textit{disrecommending} our initial guess (recall that $\vec{\theta} = \Theta \cdot \text{sgn}[\vec{\psi}]$). Because for this particular instance our initial guess is favourable, if we drive the protocol such that we disrecommend our guess, we observe a gap closing, as it is reflected in Fig.~\ref{fig:E_ov_oas_rothi}.a with $\Theta=-0.6\Omega$. On the other hand, when $\Theta=0.6\Omega$ and $\Theta=\Omega$ the initially guessed state is recommended, and the spectrum presents a clear widening of the distance between the GS and the first excited state along the last two thirds of the anneal in comparison to the direct anneal case $\Theta=0$. The effect this has on the computation is made apparent in Fig.~\ref{fig:E_ov_oas_rothi}.b, where the optimal adiabatic schedule resulting from Eq.~\eqref{svst} is presented. The case $\Theta = -0.6 \Omega$ has been left out of the plot because it results in an infinite total adiabatic time (since the gap closes), and instead we compare the recommended scenarios to the direct case. Here, the latter has resulted in a total adiabatic time $\Tilde{T}_{ad} = \int_0^1 T_{ad}(s) ds$ of  $\Tilde{T}_{ad} = 785$, while with our protocol we reach $\Tilde{T}_{ad} = 190$ for $\Theta = \Omega$. This representation evidences the great impact the enlargement of the gap caused by our technique may have in the total annealing time because of the inverse quadratic dependence.

While for this instance the differences shown in Fig.~\ref{fig:E_ov_oas_rothi}.a between $\Theta=0.6 \Omega$ and $\Theta=\Omega$  are not so obvious, the overall widening of the gap is always higher for $\Theta = \Omega$ (for a fully correct guess), in accordance to what is shown in Fig.~\ref{fig:pert_th_vsN}. This difference is captured by the optimal adiabatic schedule in Fig.~\ref{fig:E_ov_oas_rothi}.b and by the slightly greater robustness of $\Theta = \Omega$ shown in Fig.~\ref{fig:E_ov_oas_rothi}.c. Fig~\ref{fig:E_ov_oas_rothi}.c presents some results regarding the simulation of the annealing process in finite time, namely with $T = 15$.
This final time has been chosen with the intention of illustrating a case in which an unbiased anneal would provide a final probability around 50\% (i.e., far from the adiabatic limit $\Tilde{T}_{ad}$), which would lead to a somewhat poor and/or inefficient identification of the solution. In both cases, we follow the evolution of the system by starting in the initial GS and consecutively evolving for a short time $\delta t$ through exact diagonalisation. The final probabilities in Fig.~\ref{fig:E_ov_oas_rothi}.c are 0.47, 0.07, 0.71 and 0.74 for $\Theta = 0, -0.6\Omega, 0.6\Omega, \Omega$, respectively.

\begin{figure}[h]
    \includegraphics[width = 0.49\textwidth]{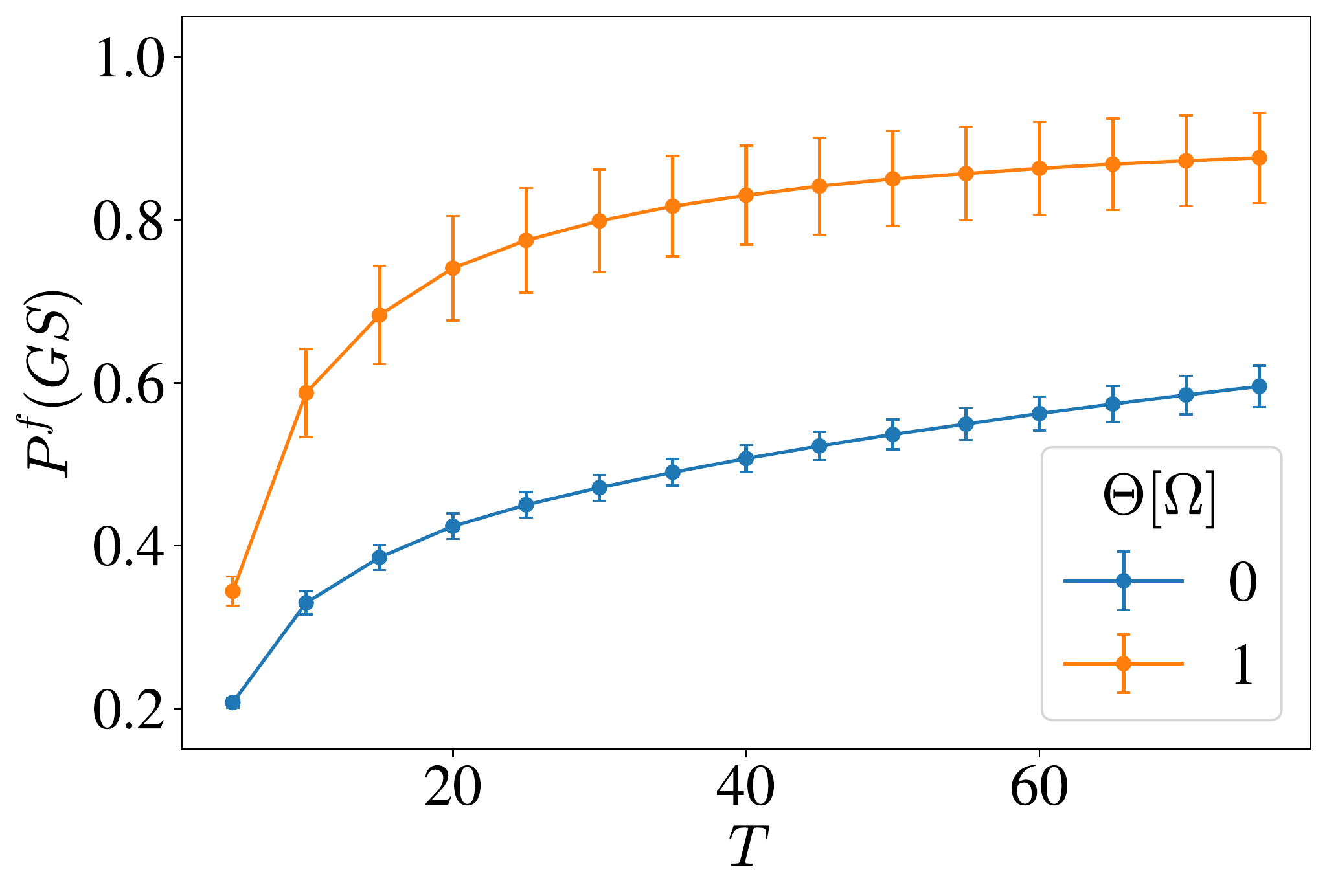}
    \caption{Final ground state probabilities for different total annealing times. The systems under consideration are of $N=8$, $h_{mean} = 0.01$ and $W=0.05$. Averages were taken over the 73 instances out of 100 where the initial guess turned out to be correct. The simulation was done using Qibo's exponential solver and adimensional time step $\delta s = dt/T = 0.01$.
    }
    \label{fig:I_Ts}
\end{figure}

In order to assess the average enhancement produced by this initial guess,  we have examined 100 random Ising instances of system size $N=8$, $h_{mean} = 0.01$ and $W=0.05$ and compared the final probability $P^f(GS)$ of obtaining the GS with our protocol. To quantify the relative improvement of the minimum gap, we define $R_{\Delta}$ as follows:
\begin{gather}
    R_{\Delta} = \frac{\min_{\Theta = \Omega} \Delta}{\min_{\Theta = 0} \Delta} - 1 
    \label{ratio_HM_Delta}
\end{gather}
We calculate $R_{\Delta}$ for the accurately guessed instances (73 out of 100), and found the typical improvement to be $R_{\Delta} = 0.96 \pm 0.41$, where we excluded 10 outlying instances for which the improvement was of one or two orders of magnitude greater. This means that, in the typical case, the minimum gap was found to be doubled for a relative guess size of $L_g /N = 1/8 = 0.125$. We have also examined $P^f(GS)$ far from adiabaticity (i.e., $T \ll \Tilde{T}_{ad} \sim 1000$ for $\Theta = 0$) in order to observe the improvement of robustness of the algorithm. Results of this study are presented in Fig.~\ref{fig:I_Ts}, where an almost doubling of the chances to find the final ground state is shown.

\subsection{3SAT} \label{sec:3SAT}

In order to probe our method in another interesting use case, we have applied it to one prototypical NP-complete problem, namely 3SAT with a unique solution (see an overview on why these are especially hard in \cite{laumann_statistical_2010}, for example). The 3SAT instances explored here have been constructed by sequentially adding randomly generated clauses and checking how many assignments still satisfy them all, until a further addition makes it impossible that all the clauses be satisfied \cite{hofmann_probing_2014}. Once obtained, these clauses can be turned into a problem Hamiltonian following the usual QUBO encoding:
\begin{gather}
    h_{c} = \left[ \frac{1}{2}(I - \sz_i) + \frac{1}{2}(I - \sz_j) + \frac{1}{2}(I - \sz_k) - I \right]^2 
    \label{SAT_clauses}\\
    H_f = \sum_{c \in \mathcal{C}} h_c
\end{gather}
where $c = c_{ijk}$ is the clause involving variables $i, j, k$ and $\mathcal{C}$ is the set containing all the clauses that make up a given instance.
\newline
As it may be noticed in \eqref{SAT_clauses}, these problems constitute an Ising model in a different parameter regime from the one studied so far, with $\sum_i h_i \approx \sum_{ij} J_{ij}$. 
Since in this case we have no clear criterion for a first guess we will construct it from the true solution, which we have obtained through exact diagonalisation.  The lack of information to make an educated guess is a consequence of our 3SAT instance being completely artificial, but in general we expect to have enough information to make some assumption based on the problem's meaning. This initial guess could also be built from an approximation provided by a classical heuristic approach.

\begin{figure}[h]
    \includegraphics[width = 0.49\textwidth]{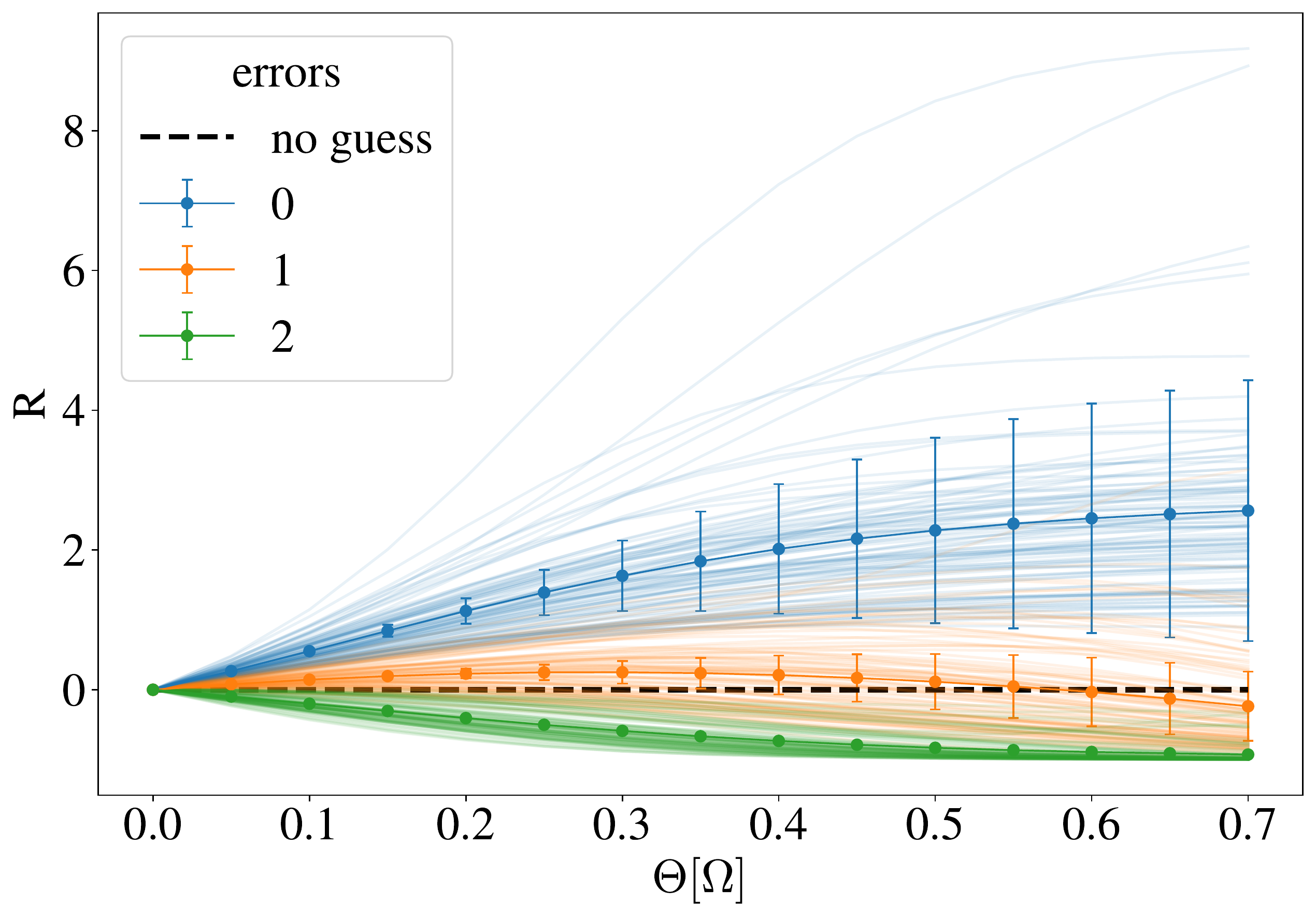}
    \caption{Improvement ratio $R$ for the probability of reaching the final ground state (see Eq.~\eqref{ratio_HM}) for 100 3SAT instances of $N=8$ spins with an initial guess of length $L_g = 3$ for different number of incorrect assignments. Solid lines with markers represent the average values and variance, while the light-coloured lines represent individual trajectories. The simulation was carried out for $T = 10$, and the maximum average improvement for the case when a single error is present was found to be $R=0.25$ at around $\Theta = 0.25 \Omega$.
    }
    \label{fig:SAT}
\end{figure}

In Fig.~\ref{fig:SAT} we present a study of the improvement of the final ground state probability obtained for a series of 3SAT instances when we consider a guess of length $L_g = 3$ for different number of incorrect assignments over the range $\Theta[\Omega] \in [0, 0.7]$. For the purpose of this assessment we have defined the improvement ratio between the final ground state probabilities $R$ as follows:
\begin{gather}
    R = \frac{P^f_{\Theta = \Omega}(GS)}{P^f_{\Theta = 0}(GS)} - 1
    \label{ratio_HM}
\end{gather}
We observe that small rotation angles are able to gain some advantage over the absence of steering even in the presence of some error in the initial assignment. Specifically, the relative improvement of the final overlap is maximised for $\Theta=0.25\Omega$, with $R_{max} = 0.25$. This aligns with the prediction of a greater robustness to error provided by choosing a small value of $\Theta$ established in Section \ref{sec:pert_GS_Hm} for the proposed methodology, despite the fact that we are now in a different parameter regime. We comment that the improvement ratio of the minimum gap $R_{\Delta}$ was also studied in this case, but since it presented very similar tendencies to the ones shown by $R$ we chose to depict the latter, as it also contains information about the enlargement of the gap throughout the whole anneal.

\section{Conclusions and outlook}

We have provided a protocol for the improvement of adiabatic annealing processes based on having reliable partial information about the sought-after solution, which is often available. This procedure does not require different interaction terms (which may be harder to implement) nor additional ancillary qubits, and enlarges the average gap throughout the whole anneal apart from widening the minimum gap. The direct consequence of this is that the adiabatic algorithm can run faster in time, and therefore be more efficient. Nevertheless, an assessment of the accuracy of the information we have about the solution is necessary in order to set an appropriate $\Theta$ that will allow us to gain something from the procedure even in the presence of incorrect assignments. 
For the Ising problems analysed in more depth here, we found a typical relative improvement of the minimum gap $R_{\Delta} \sim 1$ for a fully correct guess of relative size $L_g / N = 0.125$. We also remark that, in the case of having absolute certainty over the recommendation contained in $L_g$, this protocol will always contribute to the robustness and time efficiency of our algorithm. In addition, the proposed methodology may be used to disrecommend certain subspaces, which constitutes an interesting strategy to avoid known local minima and search for better solutions.

This work has focused solely on classical problem Hamiltonians, but this procedure may also be useful in quantum Hamiltonians where spins are precessing about the z-axis with a somewhat small angle. This last remark is necessary because we are relying on the fact that spins are localised about the z-axis in the target GS. In general, the procedure is applicable as long as the qubits (or, at least, the ones we guess over in $L_g$) are approximately localised around some known direction, since we could approach it with a suitable choice of $\vec{\theta}$ following the same intuition. This makes our protocol suitable for the improvement of quantum chemistry simulations, for example, where the level of entanglement in the ground state of the molecule of interest is often rather low. The extent to which the present results can be extended to entangled systems is an interesting venue for further research. In this same line, it would also be interesting to consider the extension of the proposed methodology to the more general context of qudits, where we should be able to benefit further from the possibility of disrecommending certain states of the individual $d$-level system. These higher-dimensional playing fields are available in superconducting platforms, for example, where further levels apart from the ground and first excited state may be addressed.

\appendix

\section{Perturbative expansion of the ground state of $H(s)$ for low $s$} \label{sec:pert_appendix}

For a more detailed analysis than the one provided in the main text, we may recall the canonical splitting of $H(s)$ (defined in Eq.~\eqref{justacatalyst}) between the unperturbed and perturbation Hamiltonians for low $s \rightarrow s^*$:
\begin{gather}
    H = \Tilde{H}_0 + \varepsilon H_f = R^{\dagger}_y(\vec{\theta}) H_0 R_y(\vec{\theta}) + \varepsilon H_f \\
    \varepsilon = \frac{s^*}{1-s^*} 
\end{gather}

We begin by the exact description of the ground and first excited states corresponding to $\Tilde{H}_0$, the unperturbed Hamiltonian, which will become our basis states:
\begin{gather}
    \ket{\varphi^-}_i = \frac{1}{\sqrt{2}} 
    (\sqrt{1 + \sin{\theta_i}} \ \ket{\uparrow}_i - \sqrt{1 - \sin{\theta_i}} \ \ket{\downarrow}_i ) \label{neg_phase} \\
    \ket{\varphi^+}_i = \frac{1}{\sqrt{2}} 
    (\sqrt{1 + \sin{\theta_i}}\  \ket{\uparrow}_i + \sqrt{1 - \sin{\theta_i}} \ \ket{\downarrow}_i )  \\
    \ket{\Phi_0} = \bigotimes_{k=1}^N \ket{\varphi^-}_k 
    \label{unperturbedGS_Hm} \\
    \ket{\Phi_1} = \frac{1}{\sqrt{N}} \sum_l \bigotimes_{k\neq l} \ket{\varphi^-}_k \ket{\varphi^+}_l \\
    \vdots \nonumber
\end{gather}

By carefully following the expressions given by second-order perturbation theory, we obtain the following description for the perturbed ground state of $H(s^*)$:
\begin{gather}
    \ket{\Tilde{\Phi}_0} = (1 + A^{(2)}_0 ) \ket{\Phi_0} + 
    (A^{(1)}_1 + A^{(2)}_1) \ket{\Phi_1} + \nonumber \\
    + (A^{(1)}_2 + A^{(2)}_2 ) \ket{\Phi_2} + A^{(2)}_3 \ket{\Phi_3} + A^{(2)}_4 \ket{\Phi_4}
    \label{Phi0}
\end{gather}
with 
\begin{gather}
    A^{(1)}_1 = \varepsilon 
    \frac{\sum_l h_l}{\epsilon\sqrt{N}} \\
    A^{(1)}_2 = \varepsilon \frac{\sum_l \sum_{m>l} J_{lm}}{\epsilon\sqrt{2N(N-1)}} \\
    A^{(2)}_0 = -\varepsilon^2 \frac{1}{2N \epsilon^2}
    \left( (\sum_l h_l)^2 + \frac{(\sum_l \sum_{m>l} J_{lm})^2}{2N(N-1)}   \right) \\
    A^{(2)}_1 = \varepsilon^2 \frac{(\sum_l h_l)(\sum_l \sum_{m>l} J_{lm})}{\sqrt{N} \epsilon^2}
    \left( 4 - \frac{1}{\sqrt{N}} \right) \\
    A^{(2)}_2 = \varepsilon^2 \left(  \frac{(\sum_l h_l)^2 \sqrt{N-1}}{\sqrt{2N}N\epsilon^2} - \frac{2 (\sum_l \sum_{m>l} J_{lm})^2}{N \epsilon^2 \sqrt{N(N-1)}}
    \right) \\
    A^{(2)}_3 = \varepsilon^2 \frac{(\sum_l h_l)(\sum_l \sum_{m>l} J_{lm}) \Gamma_3}{\epsilon^2} \left( 1 -\frac{2\sqrt{2}(N-2)}{3N\sqrt{N(N-1)}}
    \right) \\
    A^{(2)}_4 = \varepsilon^2 \frac{3\Gamma_4 (\sum_l \sum_{m>l} J_{lm})^2 }{\epsilon^2}
\end{gather}
where $\epsilon$ is the distance between levels in the unperturbed Hamiltonian (i.e., in our case $\epsilon = 1$) and $\Gamma_3, \Gamma_4$ are the relevant combinatorial coefficients that count the degrees of freedom in the 3- and 4-dimensional generalisations of an anti-symmetric matrix.

\begin{figure}[h]
    \includegraphics[width = 0.49\textwidth]{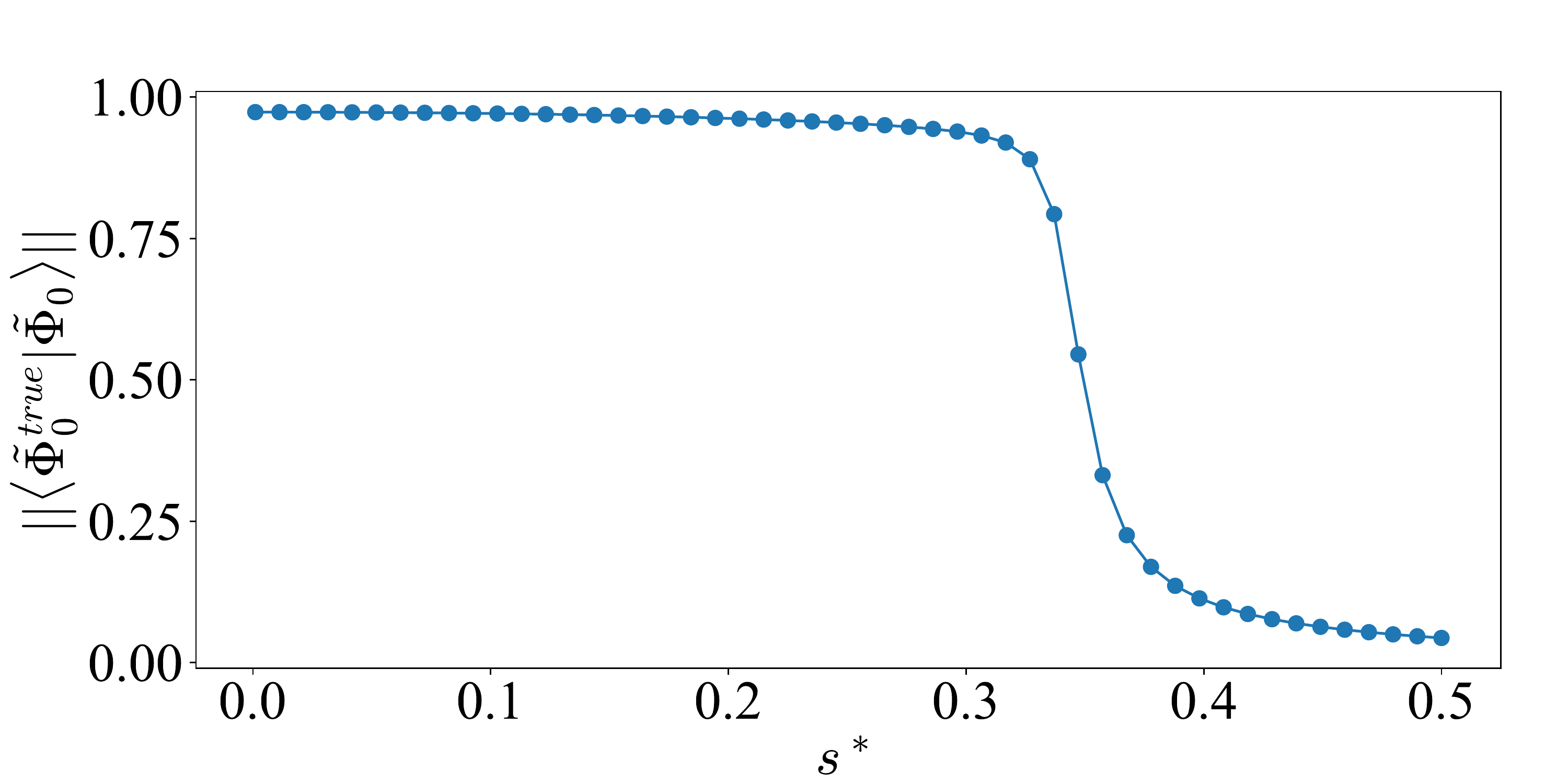}
    \caption{Overlap between the approximated $\ket{\Tilde{\Phi}_0}$ and true $\ket{\Tilde{\Phi}^{true}_0}$ wave function of the GS of $H(s^*)$ for an Ising model of $N=8$ spins, $h=0.01$ and $W=0.05$ for increasing values of $s^*$.}
    \label{fig:check_validity_expansion}
\end{figure}

With this expressions at hand, we may now study the convergence of this second-order approximation with respect to $s^*$, the point of the anneal in which we have located $H$, which will determine the relative size of the perturbation. Results of this analysis are shown in Fig.~\ref{fig:check_validity_expansion}, where we show the overlap between the ground state of $H$, $\ket{\Tilde{\Phi}^{true}_0}$, obtained numerically via exact diagonalisation, and $\ket{\Tilde{\Phi}_0}$ for an Ising model in the spin-glassy regime described by \eqref{Phi0}. We observe that for $s^* \lesssim 0.3$ our approximation holds.

\section{Scaling with the number of correct assignments in a given guess} \label{sec:scaling_correctness}

\begin{figure}[h!]
    \includegraphics[width = 0.48\textwidth]{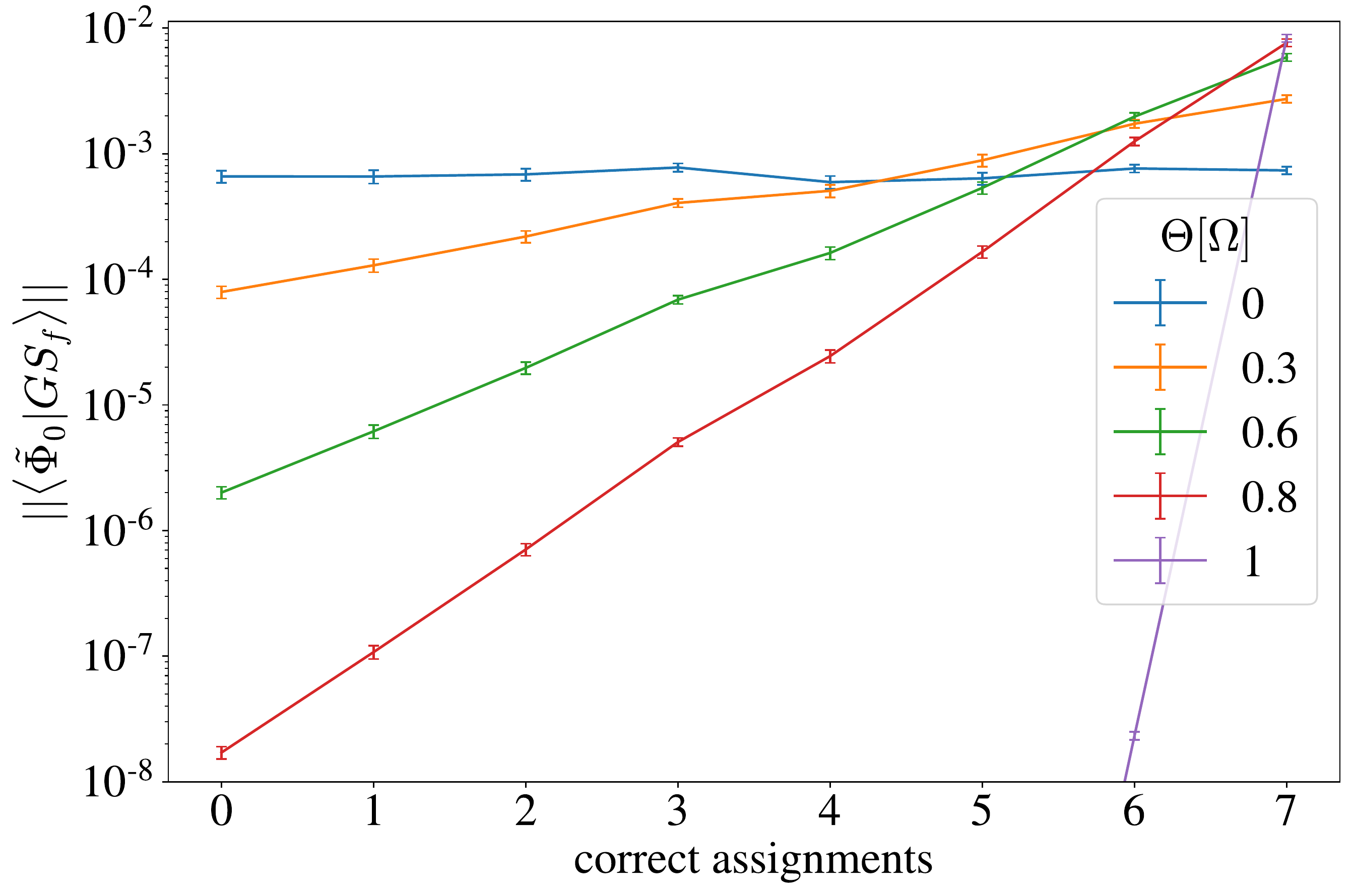}
    \caption{Overlap of the analytically approximated ground state of $H$ ($s^* = 0.3$) with the final target state depending on the number of correct assignments for different values of $\Theta$. The considered guess length is $L_g = 7$ in a system of size $N=35$. Results are averaged over 20 Ising instances of $\overline{J} \simeq -0.007, \sigma_J \simeq 14$ and $\overline{h} \simeq 1.22, \sigma_h \simeq 0.08$ (see \ref{sec:pert_GS_Hm} for the reason behind the choice of these parameters).}
    \label{fig:pert_th_vscorrectness}
\end{figure}

We also present here the analysis of the scaling of the overlap of the analytically approximated ground state $\ket{\Tilde{\Phi}_0}$ with the final ground state $\ket{GS_f}$ as a function of the number of correct assignments present in a guess of fixed length $L_g$ (for a fixed system size $N$). Fig.~\ref{fig:pert_th_vscorrectness} evidences the greater robustness to errors in the initial assignment of the lower fractions of $\Omega$. In particular, $\Theta = \Omega$ will always be detrimental unless all $L_g$ guessed qubits are correct, since any mistake is forcing the system to incur in error.

\begin{acknowledgments}
We thank M. Werner for valuable discussions. This work was supported by European Commission FET-Open project AVaQus (GA 899561), the Agencia de Gestió d'Ajuts Universitaris i de Recerca through the DI grant (No. DI74) and the Spanish Ministry of Science and Innovation through the DI grant (No. DIN2020-011168).
\end{acknowledgments}


\bibliography{references}

\end{document}